\documentclass[conference]{IEEEtran}
\IEEEoverridecommandlockouts
\usepackage{cite}
\usepackage{amsmath,amssymb,amsfonts}
\usepackage{algorithmic}
\usepackage{graphicx}
\usepackage{textcomp}
\usepackage[dvipsnames]{xcolor}
\usepackage{microtype}
\usepackage{ctable}
\usepackage{multirow}
\usepackage{url}

\usepackage{siunitx} 

\usepackage{csquotes}
\usepackage[american]{babel}

\usepackage{subcaption}
\DeclareCaptionLabelSeparator{periodspace}{.\quad}
\usepackage{pifont}
\usepackage{listings}

\definecolor{mGreen}{rgb}{0,0.6,0}
\definecolor{mGray}{rgb}{0.5,0.5,0.5}
\definecolor{mPurple}{rgb}{0.58,0,0.82}
\definecolor{backgroundColour}{rgb}{0.95,0.95,0.92}

\lstdefinestyle{CStyle}{
    backgroundcolor=\color{white},   
    commentstyle=\color{mGreen},
    keywordstyle=\color{magenta},
    numberstyle=\tiny\color{mGray},
    stringstyle=\color{mPurple},
    basicstyle=\ttfamily\footnotesize,
    breakatwhitespace=false,         
    breaklines=true,                 
    captionpos=b,   
    morekeywords={global,kernel},
    keepspaces=true,                 
    numbers=left,      
    xleftmargin=1em,
    numbersep=5pt,                  
    showspaces=false,                
    showstringspaces=false,
    showtabs=false,                  
    tabsize=2,
    language=C
}

\def\BibTeX{{\rm B\kern-.05em{\sc i\kern-.025em b}\kern-.08em
    T\kern-.1667em\lower.7ex\hbox{E}\kern-.125emX}}
    
\begin{document}
\bstctlcite{IEEEexample:BSTcontrol}

\title{Analytical Model of Memory-Bound Applications Compiled with High Level Synthesis}

\author{\IEEEauthorblockN{Maria A. D\'avila-Guzm\'an\IEEEauthorrefmark{1},
Rub\'en Gran Tejero\IEEEauthorrefmark{2}, Mar\'ia Villarroya-Gaud\'o\IEEEauthorrefmark{3} and
Dar\'io Su\'arez Gracia\IEEEauthorrefmark{4}}
\IEEEauthorblockA{DIIS-I3A, Universidad de Zaragoza --- HiPEAC Network of Excellence\\ e-mail: \IEEEauthorrefmark{1}angelicadg@unizar.es,
\IEEEauthorrefmark{2}rgran@unizar.es,
\IEEEauthorrefmark{3}mvg@unizar.es,
\IEEEauthorrefmark{4}dario@unizar.es}}


\maketitle

\thispagestyle{plain} 
\pagestyle{plain}    

\begin{abstract}

The increasing demand of dedicated accelerators to improve energy efficiency and performance has highlighted FPGAs as a promising option to deliver both. However, programming FPGAs in hardware description languages requires long time and effort to achieve optimal results, which discourages many programmers from adopting this technology.  

High Level Synthesis tools improve the accessibility to 
FPGAs, but the optimization process is still time expensive due to the large compilation time, between minutes and days, required to generate a single bitstream. Whereas placing and routing take most of this time, the RTL pipeline and memory organization are known in seconds. This early information about the organization of the upcoming bitstream is enough to provide an accurate and fast performance model.

This paper presents a performance analytical model for HLS designs focused on memory bound applications. With a careful analysis of the generated memory architecture and DRAM organization, the model predicts the execution time with a maximum error of 9.2\% for a set of representative applications. Compared with previous works, our predictions reduce on average at least $2\times$ the estimation error.

\end{abstract}




\section{Introduction}

The promise of faster and more energy efficient systems with the inclusion of heterogeneity faces several challenges, specially for systems including re-programmable hardware such as FPGAs~\cite{Waidyasooriya2017}. Fortunately for programmers, the development of High Level Synthesis (HLS) improves programmability and productivity, as CPU and GPU languages such C, C++, or OpenCL can be used to describe FPGA hardware~\cite{Gautier2016,Liang2018}.

Although HLS tools simplify programming, generating highly tuned code remains a challenge for several reasons. First, CPU and GPU optimization techniques are not always directly 
suitable for FPGA, and, second, bitstream generation takes a long time, preventing any ``trial-and-error'' optimization process. To address this issue, programmers can follow two alternatives. Either they write well-known code patterns from previous explorations~\cite{Verma2016}, or they rely on pre-synthesis analytical models for estimating performance~\cite{Zohouri2016,Wang2016,Choi2017,Liang2018}. These models analyze the RTL code generated by the HLS tools, the high-level code, or both, and often they require to instrument and run the code to obtain dynamic profiling information.


The High Performance Computing, HPC, domain represents an opportunity for HLS and FPGAs because HPC requirements include performance and energy efficiency,
and, ideally, to reuse as much code as possible. Many HPC applications are memory-bound, which together with the HLS generated code, enables to propose simple yet effective analytical models to predict their execution time when running on FPGAs. Besides, previous models focus more on the compute part, or kernel pipeline, covering the Global Memory Interconnect (GMI) connecting the kernel pipeline with the off-chip DRAM memory, in less detail. For example, the error of two state-of-the-art analytical models~\cite{Wang2016,Choi2017} multiplies by 3 when the DRAM changes and can be larger than 50\% for accesses with data dependencies. In future systems, those errors could become more prevalent because DRAM scaling is slow with a growing data rate of 7\% per year, compared with FPGA that grows capacity 48\% per year~\cite{Trimberger2015}, potentially increasing the number of memory bound applications.

The GMI is composed by multiple Load/Store Units (LSUs) which include coalescers to group requests into DRAM burst, and arbiters to order them. HLS tools generate the GMI architecture at an early stage in the tool flow, so combining information from the GMI and DRAM organizations, it is possible to build an analytical model that mainly requires static information. The model  can be easily plugged into existing models for memory-bound applications, or, even, integrated into HLS tools to guide optimizations.

The contributions of this work are: a) A description of the generated memory interconnect of an HLS tool, b) An open-source available analytical model that estimates the execution time of memory bound applications\footnote{The model with the experiment data are available at \url{https://anonymous.4open.science/r/db707fea-264d-46bf-a6f8-f2cdf455d8d2/}.}, c) A set of experiments showing that the error of the model is below 9\% for a set of representative applications.

The rest of the paper is organized as follows. Section~\ref{sec:GA} describes the HLS flow and the GMI architecture. Section~\ref{sec:MOD} introduces the model. Section~\ref{sec:METH} presents the methodology. Section~\ref{sec:RES} comments on the results. Section~\ref{sec:BACK} discusses the related work and Section~\ref{sec:CON} concludes.

\section{HLS Flow for FPGA}
\label{sec:GA}

Traditionally, HDL 
 was the preferred tool to program FPGA devices, slowing its adoption by average programmers. Recently, HLS has evolved to a point where programming from languages such as C or OpenCL for FPGA becomes an easier task. In fact, the explicit parallelism of OpenCL offers many opportunities to exploit the pipeline parallelism inherent to FPGAs, making OpenCL a good language for FPGAs.

\begin{figure}[!htbp]
\centering
\includegraphics[width=0.99\columnwidth]{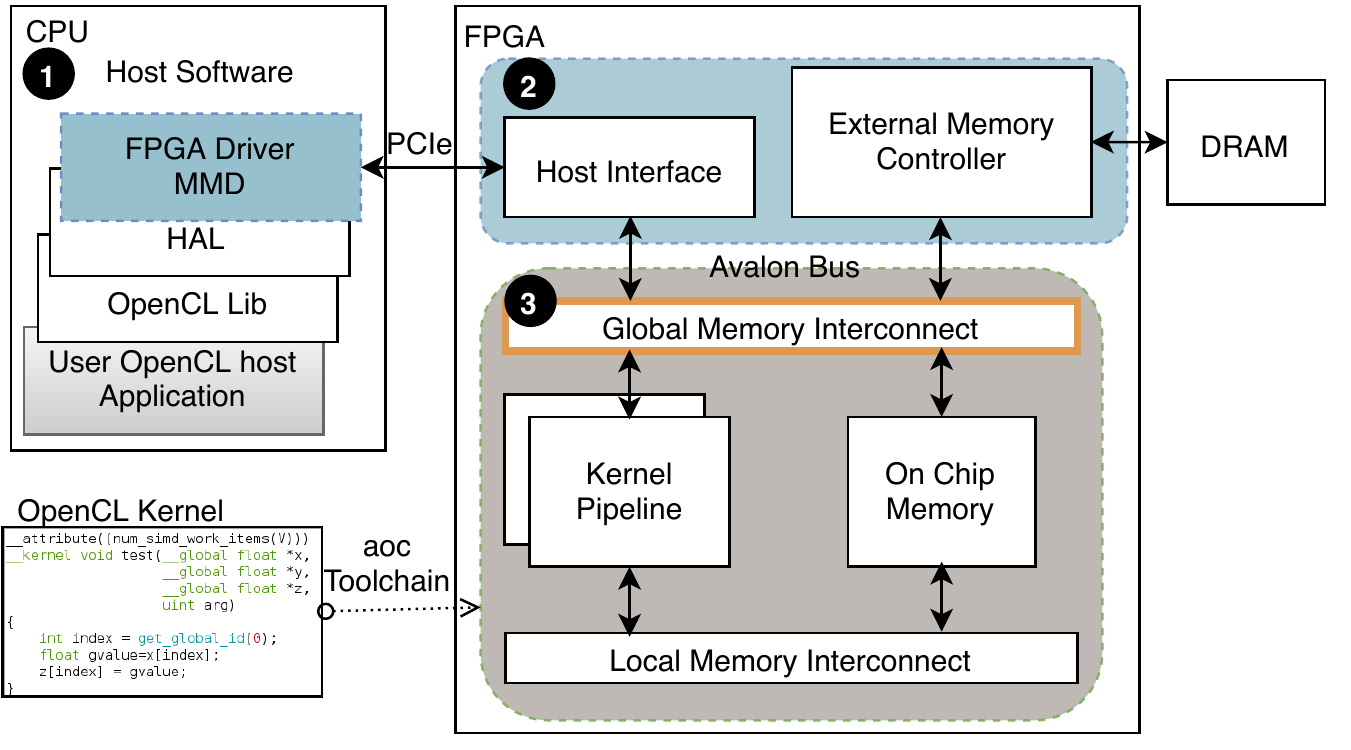}
\caption{OpenCL FPGA main elements. Blue and brown colors represent the BSP and kernel logic, respectively.}
\label{fig:BSP_arch}
\end{figure}

Figure~\ref{fig:BSP_arch} shows the main components of an OpenCL application compiled for an FPGA with the Intel OpenCL FPGA SDK. Without loss of generality, this flow is also representative for other toolchains. On the host side, \ding{182}, the application communicates with the FPGA device through 
the Board Support Package\footnote{Manufacturers often provide BSP, but advanced users can tune and re-implement them.} (BSP, in blue on the figure). The BSP implements the lower layers of the application stack such as the Memory-Mapped-Device (MMD) library, performing the basic I/O with the board and the PCIe communications. On the FPGA side, the BSP, \ding{183}, provides support to communicate back with the host and with the device memory, DRAM \ldots

From a programmer's perspective, the most important part is the kernel logic (in brown), \ding{184}, which mainly corresponds to the compiled OpenCL kernel. In fact, 
programmers seldom need to generate a new BSP\footnote{Please also note that HLS tools would always require a BSP to compile an OpenCL kernel for supporting the aforementioned low-level tasks.}. The compilation process consists of two main steps. First, a translator generates HDL code from the OpenCL, and then, a synthesis tool generates
the bit-stream. The translator creates 4 blocks from the code: Local Memory Interconnect, On Chip Memory, Kernel Pipeline, and Global Memory Interconnect, being the last two the most critical from a performance point of view.

\subsection{Kernel Pipeline}
\label{ssec:kernel_pipe}

The kernel pipeline implements the whole data and control operations. The high-level OpenCL statements are translated into a graph where each node performs an operation. To receive and send data, there are nodes that interconnects the pipeline with either the local or global memory. To exploit work-item parallelism, HLS tools implement very deep pipelines
. Splitting up the processing into small pipeline stages also helps reaching high frequency, which is another key parameter for kernel performance besides pipeline length and initiation interval
.

\subsection{Global Memory Interconnect}
\label{sec:mem}

The 
 GMI communicates the Kernel Pipeline and the main DRAM memory. In OpenCL source code, each reference to a variable hold in the global memory constitutes a \emph{global access}. Since global accesses are the main source of kernel stalls, the GMI implements several strategies to maximize DRAM throughput and kernel pipeline flow. 
Internally, the GMI architecture, as other hardware memory interfaces from Intel~\cite{IntelEMI_IP2019}, has two main components: 
LSUs, which track in flight memory operations, and arbiters. There are two independent round-robin arbiters ordering read and write accesses.


\lstset{style=CStyle}
\ctable[caption={GMI LSU Types and their modifiers for Intel FPGA SDK, the codes snippets are from Intel FPGA Guides \cite{IntelQuartusPrime19}.}, 
 star, pos=!ht,label=tab:LSUTypes,doinside=\small]
{llcccl}
{\tnote[a]{Each code snippet corresponds to line 10 in listing~\ref{code:patterns}.}
\tnote[b]{The burst-coalesced type has four modifiers affecting its organization.}}
{
\FL
LSU Type    & Description       & Pipelined & Burst & Atomic  & Code Snippets\tmark[a]
\ML
Burst-Coalesced\tmark[b] & Request are group into a set of DRAM burst         \\ 
                 
    \quad Aligned        & Index is contiguous and aligned to page size                             & \ding{52} &\ding{52}  &\ding{56}&
                \lstinline$ out = x[i];$\\
    \quad Non\_Aligned  &  Index has a modifier non-aligned to page size                         & \ding{52} &\ding{52}  &\ding{56}&
                \lstinline$ out = x[3*i+1];$ \\
    \quad Write\_ACK    &   Index to access has dependencies                  & \ding{52} &\ding{52}  &\ding{56}&
                \lstinline$ out = x[j];$\\
     \quad Cache        &    Index have repetitive dependencies             & \ding{52} &\ding{52}  &\ding{56}&   
                \lstinline$for (uint k=0; k<N; k++)$ \\
                &&&&& ~~\lstinline$z[N*i+k] = x[k];$ 
\ML
Prefetching &    Compiled as Burst-Coalesced Aligned        &  \ding{56} & \ding{52} & \ding{56} & \ML
Constant-Pipelined &   Read from a constant cache    &  \ding{52} & \ding{56} & \ding{56} &  \lstinline$z[i] = cn[i];$  \ML  
Pipelined &   Requests are submitted immediately     &  \ding{52} & \ding{56} & \ding{56} &  \lstinline$ out= lmem[li - i];$  \\
\quad Never-Stall        & Connects the LSU without arbitration                            & \ding{52} &\ding{56}  &\ding{56}&
                \lstinline$ lmem[li] = x[i];$
\ML  
Atomic-Pipelined   &   For atomic operations          &  \ding{52} & \ding{56} & \ding{52} & \lstinline$atomic\_add(\&x[0], 1);$\NN
\FL
}

Depending on the access pattern, Intel FPGA SDK~\cite{IntelQuartusPrime19} has defined 5 LSU types: two for the Local Memory Interconnect (Constant-Pipelined and Pipelined) and the rest for the GMI: Burst-Coalesced, Prefeching, and Atomic-Pipelined. To understand the access pattern of each LSU, Listing~\ref{code:patterns} and Table~\ref{tab:LSUTypes} show the code that generates them and their main features; namely, 1) Pipeline, when an LSU can support multiple active requests at a time, 2) Burst, when requests are grouped before being sent to DRAM, and 3) Atomic, which serializes the operation and guarantees atomicity. Please note that each one of these LSU features expose an increasing hardware complexity. Each \emph{global access (GA)} stated in the source code can be translated into one or several LSUs, as  Section~\ref{sec:MOD} describes.

\begin{figure*}[!ht]
\centering
\includegraphics[width=.8\textwidth]{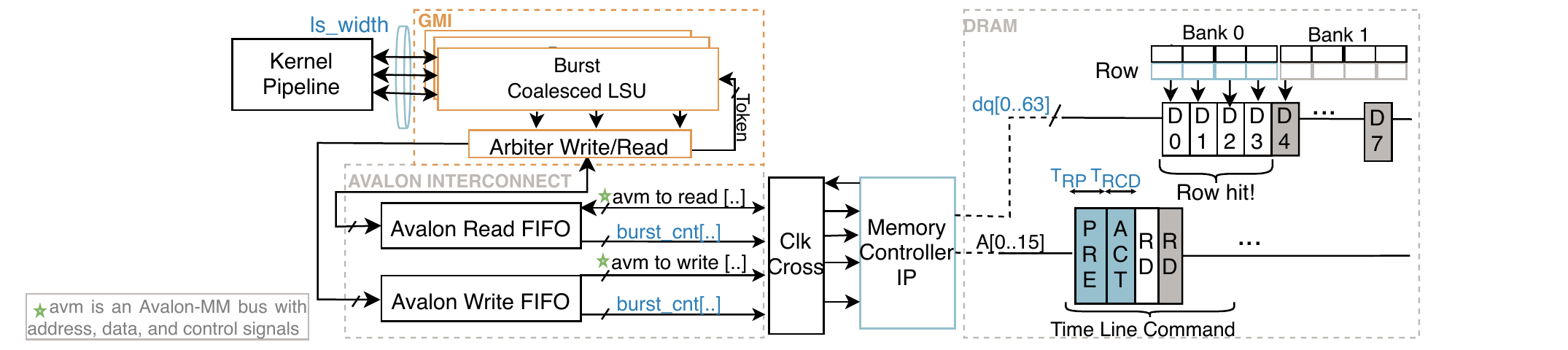}
\caption{Simplified model of a read operation with LSU Burst-Coalesced modifier. The parameter names in blue are used in the model.}
\label{fig:GMI_LSU}
\end{figure*}

Each LSU type provides a different maximum bandwidth, being the burst-coalesced (BC) with aligned modifier the most efficient type because it maximizes DRAM effective-utilization. Figure~\ref{fig:GMI_LSU} shows a read operation generated by a BC LSU. Each LSU has a coalescer unit that tries to group continuous memory address into a single burst DRAM operation. Eventually, the read arbiter dispatches this operation to the FIFO into the Avalon Interconnect in order to issue a DRAM access to the Memory Controller IP through the Avalon Bus.
The benefits come from the DRAM organization~\cite{Zheng2010} because during a read operation at least 3 commands are required: precharge (PRE), activate (ACT), read out (RD). PRE opens a row in every bank, then, ACT opens a row in a particular bank, and RD read the burst out back to the controller. When an LSU receives a requested address, it attempts to group consecutive addresses into a burst, the \emph{burst\_cnt} bus size defines the maximum number of burst requests at compilation time, because contiguous access to memory enables to hide the overhead of PRE/ACT commands.

\begin{lstlisting} [style=CStyle, caption={OpenCL Code for access patterns in Table \ref{tab:LSUTypes}}
,label={code:patterns},floatplacement=T, float=t]
#define N 1024
int random_vector[N]={5,1023, 450, 100, ...}
__kernel void 
test_patterns(  global int *restrict x,
                global int *restrict z,
                constant int *cn )
{   int i   = get_global_id(0); 
    int j   = random_vector[i];
    int out = 0;  local int lmem[1024];
    //Code Snippet form Table I
    z[0]    = out;
}
\end{lstlisting}


In burst coalesced LSU, three limits trigger a request to the DRAM: 1) Burst\_cnt bus, that usually corresponds to memory page size, 2) Maximum number of threads 
allowed to be coalesced
and 3) Time out to minimize stalls in the kernel pipeline when the consecutive requests can not be coalesced. The compiler can modify this LSU depending on the memory access pattern and other attributes \cite{IntelQuartusPrime19}; e.g., in the case of data dependencies, the compiler infers a Write-ACK LSU with a work-item level coalescer.


In a \textit{Prefetching} LSU, the behavior is similar to the burst coalesced, but anticipating 
a large amount of data. 
For write operations, it uses a burst coalesced non-aligned LSU structure. In highend FPGAs, the compiler generates a Burst Coalesced LSU with given Intel SDK 
code. 

The last LSU for GMI is the \emph{atomic-pipeline}, Intel provides limited support only for 32-bit integers and it is considered one of the most expensive functions in HLS.

\subsection{Performance Estimation for FPGA}

Both the Kernel Pipeline and the GMI have a huge impact on performance. Kernel performance has been modeled to predict the execution time aiming at the automatization of the compilation process~\cite{Liang2018,Wang2016,Choi2017}, but the memory component has been simplified, losing details that can expand the search space to optimize. This simplification, valid for old FPGA models, does not apply for newer models because the  kernel resources' number has grown faster than the external memory; e.g., an Intel Stratix 10 reaches 9 TFLOPS while the newer Intel Agilex reaches 20 TFLOPS, although DRAM has improved from DDR4 @ 1333 MHz / 2666 Mbps to DDR4 @ 1600 MHz / 3200 Mbps or DDR5 @ 2100 MHz / 4400 Mbps, namely, kernel performance has doubled, meanwhile the memory bandwidth has not~\cite{IntelStratix,IntelAgilex}. Comparing the on-chip memory and the external memory, the external memory is more critical because the throughput of the DRAM is 380$\times$ worst than that of the embedded memory, and its size is 80$\times$ larger~\cite{IntelStratix}. Therefore, a more accurate model of memory becomes desirable to estimate performance for the latest FPGA models.
\section{Analytical Model}
\label{sec:MOD}


In the kernel source code, a \emph{global access (GA)} is translated into one or several LSUs at the GMI. 
The compiler determines the proper type of LSU for each \emph{GA} according to a static analysis. As described in Section~\ref{sec:GA}, LSUs are part of the GMI that is modelled in this section. This model just relies on information available up to the translation phase (OpenCL to Verilog) providing accurate execution-time estimations for memory bound kernels without the long 
delays of the full compilation process. 



Table \ref{tab:parameters} summarizes the input parameters to the model with their respective sources which are described below:
\begin{enumerate}
    \item Report: Html file generated in intermediate compilation using \emph{aocl -rtl}. It shows the kernel basic blocks and the types of the LSUs.
    \item Verilog: These files instantiate the parameters of IPs and show stop conditions of memory controller. These files are generated with \emph{aocl -rtl}. 
    \item Users: For dynamic loops, users are required to provide the iteration limit, since it is not available at compile time. Please note that the user information could be automatically inferred by a compiler pass.
    \item Datasheets: The DRAM datasheets  provide the timing and the organization of DRAM memory chips.
\end{enumerate}

\ctable[
caption={Variable descriptions, if possible, names come from the HLS tools }, 
label=tab:parameters, doinside=\small
]{lll}{
\tnote[n]{For every $i$ LSU in a maximum of $\#lsu$} 
}{
\FL
Variable & Definition & Source 
\ML
$\#lsu$         & Number of load/store units            & Report\NN
$ls\_width^{i}$   & Memory width of $i$ LSU [bytes]      & Report

\ML
$burst\_cnt^{i}$  & Size of Avalon $burst\_count$ port  & Verilog  \\
                &  param:BURSTCOUNT\_WIDTH              &   
                \NN
$max\_th^{i}$   & Maximum threads in a burst            & Verilog  \\
                & param:MAX\_THREADS                  &   

\ML
$\delta$        & Address stride of memory access       & User \NN
$ls\_acc^{i}$   & Number of access of each LSU          & User \NN
$ls\_bytes^{i}$ & Bytes of a single $ls\_acc$           & User \NN
$f$             & Kernel vectorization $SIMD \cdot Unroll$      & User
\ML

$dq$            & Memory data width [bytes]            & Datasheet\NN
$bl$            & Memory burst length                   & Datasheet\NN
$f\_mem$        & Memory frequency [Hz]                 & Datasheet\NN
$T_{RCD}$       & Row activation time [s]                & Datasheet  \NN
$T_{RP}$        & Precharge row miss time [s]            & Datasheet\NN
$T_{WR}$       & Time to recovery from Write[s]          & Datasheet
\LL
}


To begin with, let $T_{exe}$ be the execution time estimated as the sum of minimum time, $T_{ideal}$, of every transaction from every LSU plus an overhead time, $T_{ovh}$, which depends on the LSU type, as shows  Eq.~\ref{eq:T_total}. 

\begin{equation}
T_{exe}= \sum_{i=1}^{\#lsu} \delta^{i} \cdot (T_{ideal}^{i} + T_{ovh}^{i})
\label{eq:T_total}
\end{equation}

$T_{ideal}^{i}$ only depends on maximum memory data transfer, and, hence, is the same for all LSU types, while $T_{ovh}^{i}$ changes; e.g., when a memory access has a stride of $\delta^{i}$, then coalescing LSUs do not use all data burst, which increases the amount of memory transactions. In other words, $T_{ideal}^{i}$ corresponds to the minimum time for bringing every data which is the amount of bytes read by an LSU divided by the DRAM bandwidth, as shown in Eq.~\ref{eq:t_min}, with $\small{bw\_mem= dq \cdot 2 \cdot f\_mem}$ (The 2 corresponds to DRAM's double-rate).


\begin{equation}
T_{ideal}^{i}=\frac{ ls\_bytes^{i} \cdot ls\_acc^{i}  }{bw\_{mem}}  
\label{eq:t_min}
\end{equation}

\subsection{Burst-Coalesced LSU}

In a memory bound application, the Avalon FIFO needs to be full of requests to maintain the DRAM with high occupancy. When the kernel pipeline does not make enough requests to fill the memory burst request before time out, the memory will fall down in low occupancy. In this state, the kernel is compute bound which has already been covered in previous works \cite{Wang2016,Choi2017}.

Another cause of low memory occupancy is when the $ls\_width$ does not have enough data to allow the maximum burst size because $ls\_width$ depends on the vectorization factor ($f$) as well. To evaluate this condition, the relation between the number of kernel request to the LSU ($ls\_width$) and the amount of data that DRAM can fulfil ($dq \cdot bl$) is calculated as shown in Eq.~\ref{eq:underused}. Let $K_{lsu}^{i}$ be the influence of each type of LSU modifier. When this condition is fulfilled, the kernel is defined as memory bound.


\begin{equation}
\parbox{4em}{Kernel Bound}\Rightarrow
\begin{cases}
\parbox{6em}{Memory high-occupancy}
& \sum_{i=1}^{\#lsu}  \frac{ls\_width^{i}}{dq \cdot bl \cdot K_{lsu}^{i}}\geq 1
\\
\text{Compute} & \text{otherwise}

\end{cases}
\label{eq:underused}
\end{equation}


Once the kernel is defined as memory bound, $T_{exe}$ can be calculated with Eq.~\ref{eq:T_total}. To achieve $T_{ideal}$ is possible for contiguous memory addresses, this type of access hides PRE/ACT latencies, as was shown in Fig.~\ref{fig:GMI_LSU}. Furthermore, bank-interleaving memory controller can completely hide opening new banks~\cite{IntelEMIF2019} until the $\#lsu$ is less than two. When the $\#lsu$ increases, this forces the DRAM to open a new row,  adding a time overhead $T_{ovh}$.

The $T_{ovh}^{i}$ will be proportional to DRAM latency of opening a new page, given by row miss commands ($T_{row}$). These can be calculated by the amount of times that an $i$ LSU have to open a new row which depends of the amount of burst transactions, with a size of $burst\_size$, required to request the total bytes ($ls\_acc \cdot ls\_bytes$), formulated as in in Eq.~\ref{eq:T_ovh}. It should be noted that latency of LSU and the amount of data in a the Avalon FIFO would hide the kernel latency, for this reason only the DRAM latency is considered.


\begin{equation}
T_{ovh}^{i}=
\begin{cases}
0                              & \#lsu < 2
\\
\frac{ls\_acc^{i} \cdot ls\_bytes^{i}}{burst\_size^{i}} \cdot T_{row} & \text{otherwise} 
\end{cases}
\label{eq:T_ovh}
\end{equation}

The $burst\_size$, $T_{row}$ and $K_{lsu}$ estimations for each LSU modifier are analyzed  in the subsections \ref{aligned} to \ref{write-ack}.

\subsubsection{Burst Coalesced Aligned LSU}
\label{aligned}

This modifier is generated when all the kernel requests are contiguous memory addresses aligned to page size. 

Two kind of IPs have been detected in this type of access, one is the ``simple" whether the program only has one LSU and every kernel request is sent directly to the memory controller, without FIFO registers. This simplifies the hardware and maximizes the memory throughput. With more LSUs the complete architecture is generated as was shown in Fig.~\ref{fig:GMI_LSU}

To calculate the memory constrain, the value of $K_{lsu}=\delta$  means that one burst operation is executed per cycle, but it is limited by strides.

The next step is the estimation of the DRAM $burst\_size$. DRAM sets the minimum burst transaction size to $dq \cdot bl$, but it can transfer multiple consecutive burst for the same
open row giving Eq.~\ref{eq:busrt_coales}.  $burst\_cnt^{i}$ represents the binary logarithm of the transaction size, as shown in Fig.~\ref{fig:GMI_LSU}).


\begin{equation}
burst\_size^{i}= 2^{burst\_cnt^{i}} \cdot dq \cdot bl
\label{eq:busrt_coales}
\end{equation}

Finally, to estimate $T_{row}$ is not trivial because the controller can overlap commands due to reordering strategies and the page policy \cite{Wang2014}, then, this model takes into account the inter-command delay for row buffer misses~\cite{Choi2017} using ACT/PRE latencies, as Eq.~\ref{eq:T_row} shows. The command sequence PRE, ACT, read and write are considered with the same minimum timing. 

\begin{equation}
T_{row}=T_{RCD} + T_{RP} 
\label{eq:T_row}
\end{equation}

\subsubsection{Burst Coalesced Non-Aligned LSU}  
\label{non-aligned}
Both Aligned and Non-Aligned LSUs try to coalesce multiple thread request in a single burst command; however, the $\delta$ stride of Non-Aligned adds a new trigger for memory request, the number of threads, $max\_th$, that have been launched and coalesced.      



Eq.~\ref{eq:limth_na} calculates this constrain, named $max\_reqs$, representing the maximum size of a DRAM burst-coalesced request. 
During the assembly of a request, two conditions can trigger the request issue towards the memory, either the amount of requested data equals a DRAM page, or the number of coalesced threads have reached $max\_th$.  

When a coalescer is assembling a request, either the request occurs when the amount of data requested is equals to a DRAM page or when the number of coalesced threads have reached $max\_th$. This limit is affected by $\delta$,
effectively reducing the effective burst request. In the other case, the $\delta$ fraction of $ls\_width$ is the effective burst size as  Eq.~\ref{eq:busrt_na} shows.  Please note that $ls\_width$ should be bounded by DRAM page size.

\begin{equation}
\label{eq:limth_na}
\begin{aligned}
max\_reqs^{i}=\frac{max\_th \cdot ls\_width^{i}}{\delta+1} 
\end{aligned}
\end{equation}

\begin{equation}
\label{eq:busrt_na}
 burst\_size^{i}=
\begin{cases}
\frac{max\_reqs^{i}}{\delta}    & max\_reqs^{i}\leq 2^{burst\_cnt^{i}}\cdot dq \cdot bl \\
\frac{ls\_width^{i}}{\delta}  & \text{otherwise}
\end{cases}
\end{equation}

Finally, the parameters $K\_lsu$ and  $T_{row}$ are the same ones as for burst coalesced aligned LSU.

\subsubsection{Burst Coalesced Write-Acknowledge LSU }                              \label{write-ack}
When data dependencies occur, the compiler generates a write acknowledgement signal to guarantee the correct ordering of accesses~\cite{IntelQuartusPrime19}. Therefore, the burst size equals the aligned case from Eq.~\ref{eq:busrt_coales}, $K\_{lsu}$ equals 1, and most important, each burst only consumes $ls\_bytes$ increasing the total time by $\frac{dq \cdot bl}{ls\_{bytes}}$. The write-ack signal adds a write command to the DRAM access, increasing the $T_{row}$ delay as Eq.~\ref{eq:T_row_wack} shows. 

\begin{equation}
T_{row}=T_{RCD} + T_{RP} + T_{WR}
\label{eq:T_row_wack}
\end{equation}

\subsection{Atomic-pipeline LSU}

Atomic-pipeline LSU only supports integer data type without burst commands; hence, its stride is always 1, $\delta=1$. Every atomic operation executes a read and a write DRAM commands. For example, \texttt{atomic\_add} from Listing~\ref{code:atomic_f} atomically sums \texttt{val} to \texttt{p}, which is atomically read and written. When \texttt{val} is constant within a loop or for multiple work items, then the compiler performs $f$ operations atomically.

\begin{lstlisting} [style=CStyle, caption={Atomic-pipeline add prototype function},label={code:atomic_f}]
int atomic_add(volatile __global int *p, int val);
\end{lstlisting}

Eq.~\ref{eq:T_ovh_atomic} shows the resulting $T_{row}^{i}$, including the two accesses, and $T_{ovh}^{i}$, depending on the $f$ factor.


\begin{equation}
\begin{aligned}
T_{row}^{i} =2\cdot(T_{RCD} + T_{RP}) + T_{WR}\\
T_{ovh}^{i} =
\begin{cases}
\frac{T_{row}^{i}}{f} & $val$~\text{is constant}\\
T_{row}^{i} & \text{otherwise}\\
\end{cases}
\label{eq:T_ovh_atomic}
\end{aligned}
\end{equation}



\section{Methodology}\label{sec:METH}

All the experiments have been run on an Intel Stratix 10 GX FGPA Development Kit with 2GB of DDR4 DRAM organized in a single DIMM and 4 memory banks. Table~\ref{tab:parameters_value} shows the required DRAM parameters for the model, running at 1866MHz. The rest of parameters come from the intermediate compilation of the
Intel FPGA SDK for OpenCL 18.1.  For burst coalesced aligned and non-aligned LSU, $\delta$ variations are validated scaling the array accesses by $\delta$. In the non aligned case, an offset argument is added to the scaled index.

To validate the model,  two types of benchmarks are analyzed: first, a set of microbenchmarks, targeting each LSU type, where  the vectorization factor $f$ and the global access ($\#ga$) number vary using the Listing~\ref{code:metodology}. 

\ctable[caption={Fixed variable value to evaluate the LSU model, all variables (Var.) are defined in Table~\ref{tab:parameters}},
label={tab:parameters_value}, doinside=\small]{lr|lr|lr}{}{
\FL
Var. & Value & Var. & Value & Var. & Value \\
\ML
$f\_mem$        & 933.3     &$dq$            & 8  &$bl$             & 8 \NN
$T_{RCD}$       & 13.5e-9      & $T_{RP}$      & 13.5e-9 & $T_{WR}$      & 15e-9 
\LL
}

\begin{lstlisting} [style=CStyle, caption={OpenCL microbenchmark for LSU Coalesced Aligned},label={code:metodology}]
__attribute((num_simd_work_items(SIMD)))
__kernel void test_coalesced(
  __global const int *restrict x0,.., xn 
  __global const int *restrict z   )
{
    int id = get_global_id(0);
    //code snippet from Table II for each LSU 
}
\end{lstlisting}

 A second validation is performed in 9 different memory bound HPC benchmarks from 
the following sources: Intel FPGA SDK, Xilinx SDAccel, Rodinia FPGA~\cite{Zohouri2016}, 
 and FBLAS~\cite{DeMatteis2019}, which input channels were modified to fit the DRAM inputs.

The execution time is measured with aocl report, which compared with host measurements can have $\sim 8\%$ consistent difference. The atomic cases are measured with OpenCL events due to atomic LSU does not have dynamic counters implemented. Finally, the proposed model is compared with two state-of-the-art works~\cite{Wang2016,Choi2017}.
\section{Model Validation}
\label{sec:RES}

Two groups of experiments check the model accuracy: microbenchmarks, to dig into the keys of the model by sweeping the most common parameters such as $SIMD$ vector lanes, $\delta$, or $\#lsu$, varying the number of global access (\#ga), and, the second group being applications to showcase its effectiveness in real scenarios. 

\begin{figure*}[!htbp]
\centering
\includegraphics[width=2.0\columnwidth]{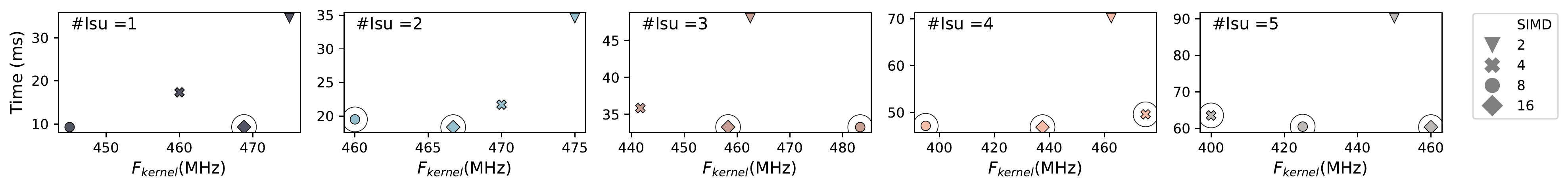}
\caption{Execution time vs. kernel frequency in a burst coalesced aligned LSU varying $\#lsu$ and $SIMD$ vector lanes. Encircled markers correspond to memory bound kernels.}
\label{fig:frq_dependency}
\end{figure*}

The model assumes that execution time depends on memory delay more than on kernel frequency for memory bound applications. To verify this claim, Fig.~\ref{fig:frq_dependency} shows the execution time for multiple sum reduction kernels with burst coalesced aligned LSUs (line 2 of Listing~\ref{code:BC}) varying $\#lsu$ and $SIMD$ vector lanes\footnote{Other LSU types produce the same results that are not shown for clarity.}. For memory bound kernels, encircled markers, $F\_{kernel}$ does not affect execution regardless $\#lsu$ and $SIMD$, because the memory delay dominates execution time. For non-memory bound kernels, uncircled markers, memory width, set by $SIMD$, affects performance more than $F\_{kernel}$ as well. Both results show the importance of early memory optimizations writing HLS code.

\subsection{Microbenchmarks}
\label{sec:validation}

For the sake of completeness, each LSU modifier is evaluated separately. The evaluation comprises the microbenchmark from Listing~\ref{code:metodology} with a body tuned to the LSU type and modifier. As shown in Listing~\ref{code:BC}, every loop body is based on sum reduction to easily change the number of GA, $\#ga$.

\begin{figure*}[!htbp]
\centering
\begin{subfigure}{2.0\columnwidth}
\includegraphics[width=1.0\columnwidth]{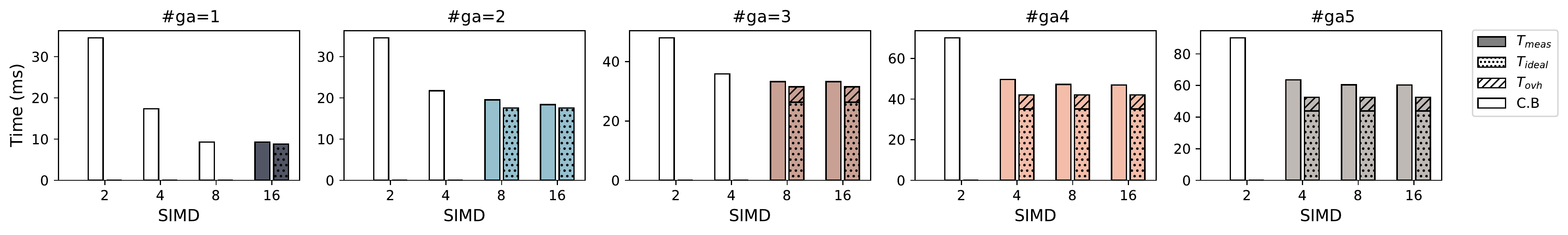}
\caption{Burst Coalesced Aligned LSU}%
\label{fig:Coalesced_BW_1866}
\end{subfigure}\hfill%

\begin{subfigure}{2.0\columnwidth}
\includegraphics[width=1.0\columnwidth]{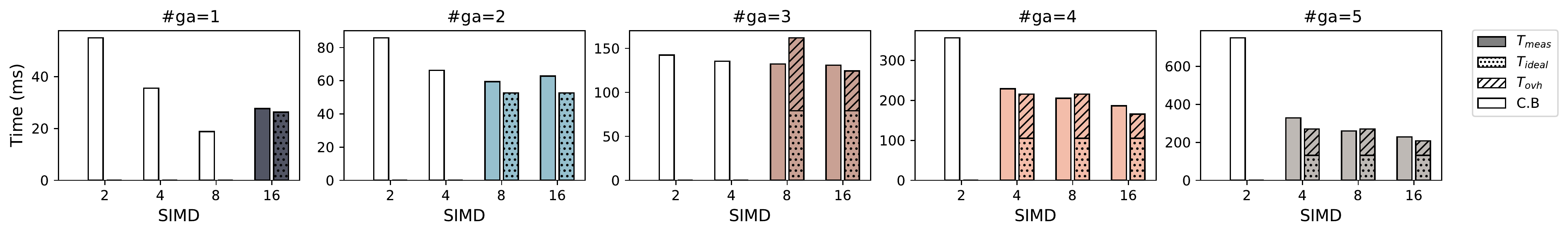}
\caption{Burst Coalesced No Aligned LSU}%
\label{fig:CoalescedNo_BW_1866}
\end{subfigure}\hfill%

\begin{subfigure}{2.0\columnwidth}
\includegraphics[width=1.0\columnwidth]{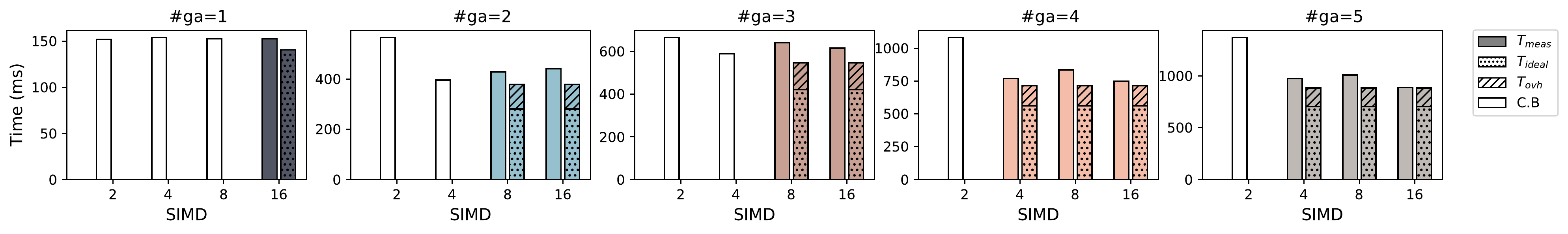}
\caption{Burst Coalesced Write Acknowledge LSU}%
\label{fig:CoalescedWACK_BW_1866}
\end{subfigure}\hfill%

\begin{subfigure}{2.0\columnwidth}
\includegraphics[width=1.0\columnwidth]{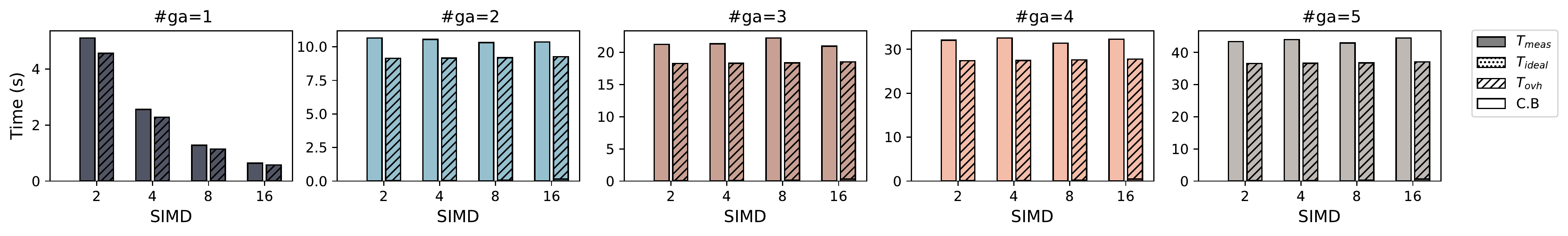}
\caption{Atomic-pipeline LSU}%
\label{fig:atomic_BW_1866}
\end{subfigure}\hfill%

\caption{Measured ($T_{Meas}$) and Estimated ($T_{ideal} + T_{ovh}$) time for all the LSU's types varying $SIMD$ vector lanes and global access ($\#ga$).
The bars with dots and lines represent  $T_{ideal}$ and $T_{ovh}$, respectively. Non memory bound kernels $(C.B)$ are detected (empty bars) and not estimated.}
\label{fig:ME}
\end{figure*}

Fig.~\ref{fig:ME} shows the measured and estimated time for every type LSU under test varying $SIMD$ and $\#ga$.
Empty bars represents non memory bound kernels, as detected with Eq.~\ref{eq:underused}. For all the combinations, error remains bellow 15.7\% for 75\% of the benchmarks with a maximum error of 27.9\%.


\begin{lstlisting} [style=CStyle, caption={Sum Reduction microbenchmark for Burst Coalesced LSUs. Each modifier is run separately},label={code:BC}]
    // Aligned
    z[id] = x1[id] + ... + xn[id];
    
     // Non-Aligned
    z[3*id+1] = x1[3*id+1] + ... + xn[3*id+1];
    
    // Write-Acknowledge
    int id = rand[i]; //work item index
    z[id] = x1[id] + ... + xn[id];
\end{lstlisting}


\begin{figure}[!htbp]
\centering
\begin{subfigure}{0.48\columnwidth}
\includegraphics[width=1.0\columnwidth]{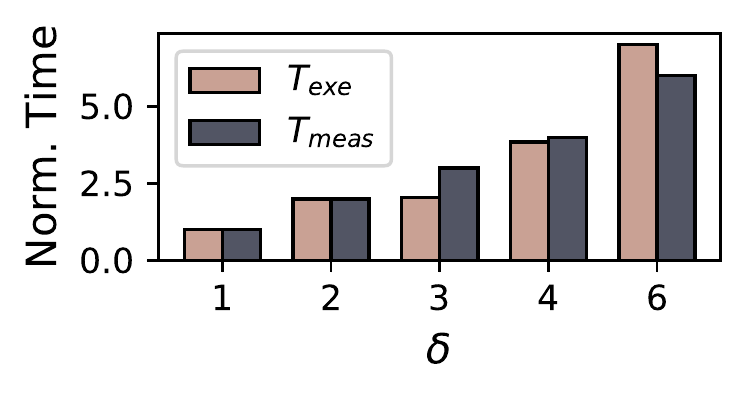}
\caption{Burst Coalesced Aligned LSU}%
\label{fig:Coalesced_BW_1866_delta}
\end{subfigure}\hfill%
~
\begin{subfigure}{0.48\columnwidth}
\includegraphics[width=1.0\columnwidth]{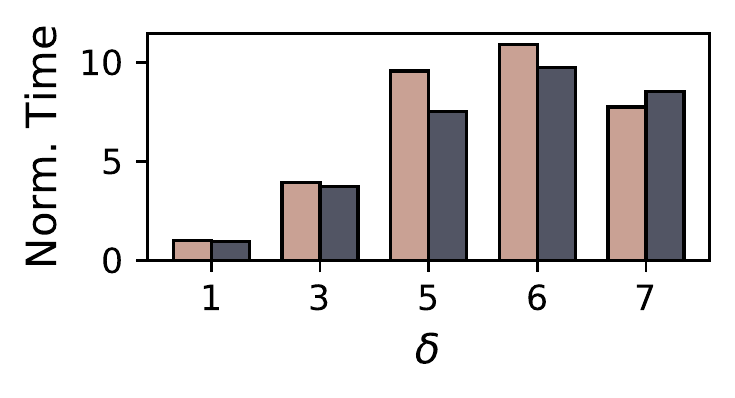}
\caption{Burst Coalesced Non aligned LSU}%
\label{fig:CoalescedNo_BW_1866_delta}
\end{subfigure}\hfill%

\caption{Measured ($T_{Meas}$) and estimated ($T_{exe}$) time are normalized to $T_{Meas}$ in $\delta=1$. The experiment varies $\delta$ with fixed values of $\#lsu=3$ and $SIMD=16$ in a) Burst coalesced aligned and b) Burst coalesced non-aligned LSUs}
\end{figure}

\subsubsection {Burst Coalesced Aligned LSU}
Digging into each type, Fig.~\ref{fig:Coalesced_BW_1866} compares the measured, $T_{meas}$ and estimated, $T_{exe}$ as the sum of $T_{ideal}$ (dots) and $T_{row}$ (lines), times for burst coalesced aligned, In this modifier, each global access generates one LSU ($\#ga$ equals to $\#lsu$).  
For all the cases, error remains below 10\%, the probable source is the simplification of the DRAM model and the refresh time, which can reduce memory efficiency around 3.5\%~\cite{IntelEMIF2019}. The experiment also evidences that the higher the $\#lsu$, the higher the $T_{ovh}$, so DRAM bandwidth reduces 26\%, from 14.2 to 10.5 GB/s. For this reason, programming strategies such as \textit{Array of Structures} reducing $\#lsu$ should be preferred.




The model shows a linear dependency to the stride parameter given by $\delta$, increasing the number of times that the controller should access to DRAM as  Fig~\ref{fig:Coalesced_BW_1866_delta} shows. Here the times are normalized to measured time with $\delta=1$ and evidence how the estimation follows the linear tendency, marked with points. Notice that this LSU can not be generated with $\delta=5$ because compiler does not detect the 
DRAM page size's alignment.



\subsubsection {Burst Coalesced Non-Aligned LSU} BCNA is depicted in line 5 of Listing~\ref{code:BC} for a $K_{lsu}=3$. Similarly to aligned modifier, in BCNA the global access is also supported by just one LSU. BCNA shows a larger error than BCA, between 4 and 21\%, because the latency of BCNA coalescer has a large variance; e.g., the number of required address comparisons depends on the coalescer state. Please also note that error does not correlate neither with $SIMD$ vector lanes nor $\#lsu$, suggesting that the model correctly tracks parameter changes.



Also, for $SIMD$ and $\#lsu$
larger than 4 and 3, respectively, the number of threads in a burst significantly impacts execution time, which increases linearly and not exponentially as $SIMD$ does.  This ``$max\_th$ effect'' can also be seen varying $\delta$ as Fig.~\ref{fig:CoalescedNo_BW_1866_delta} shows for  $SIMD=16$ and $\#lsu=3$, times are normalized to $\delta=1$. With $\delta=7$ the $max\_th$ restriction appears optimizing the access that increases with strides. In comparison with an aligned LSU, the performance is reduced in median a 60\% due address comparison increases and burst window is reduced to avoid long kernel stalls.



\subsubsection{Burst Coalesced Write-Acknowledge LSU}

The evaluation of this LSU uses  the microbenchmark in Listing~\ref{code:metodology}, with the code snippet in lines 7 to 9 of Listing~\ref{code:BC}.
  

A vector of constant values is generated by software with random values between 0 and 2048, reducing the probabilities of coalescing (2048 over 64 floats that can be coalesced in DRAM). 

The SIMD vector lanes in the previous analyzed LSU affected the $lsu\_width$, by contrast, with burst coalesced write acknowledge LSU the $lsu\_width$ remains constant. To increase the vectorization, the compiler generates so many LSU as the desired SIMD by each global access. The assumption is that every thread is accessing to different memory locations, controlling the memory consistency with the Ack signal. The Fig.~\ref{fig:CoalescedWACK_BW_1866} shows the comparison between the measured and estimated execution time.

The execution time is the worst of burst coalesced modifiers growing $24\times$ more than aligned LSU. The read operations show a stall on read until 98\% with two LSU. To optimize these cases the programmer should evaluate a balance between the data dependency with writes vs. the use of on-chip memory with a tiling strategy.

\subsubsection{Atomic-pipeline LSU}

The evaluation of this LSU uses the microbenchmark in Listing~\ref{code:metodology}, with the code snippet in Listing~\ref{code:atomic}. To generate a single global access ($\#ga=1$), $xn[id]$ has to be replaced by $id$, otherwise, each atomic operation generates its own global access to avoid coalescing.

\begin{lstlisting} [style=CStyle, caption={OpenCL microbenchmark for atomic-pipeline LSU},label={code:atomic}]
    atomic_add(z[0], x[id]);
    ...
    atomic_add(z[n], xn[id]);
\end{lstlisting}

In general, atomic-pipeline LSU does not change the $lsu\_width$, as burst coalesced does, making $T_{ovh}$
the most significant component of this LSU.
Fig~\ref{fig:atomic_BW_1866} shows that execution time increases linearly with $\#ga$ and  maximum error of 16\% corresponding to unaccounted $5 ns$ per atomic operation. This delay is near the time between the beginning of the internal write transaction and that of the following read command in the same group and same bank ($T_{WTR}$).



To quantify the LSU impact on kernel performance, we analyze read stalls. For burst coalesced aligned and non-aligned, read stalls are under 20\% because the coalescer partially hides the $\delta$ induced delay.
Meanwhile, write ACK LSU shows stalls over 50\% 
as the extra signalling serializes the requests. 
The atomic-pipeline modifier cannot be measured because profiling is unsupported, but it is safe to assume that stalls will be high due to atomicity requirements.




\subsection{Applications}

We validated the model with 9 memory bound applications, mixing single task and NDRange kernels with and without channels. Table \ref{tab:applications} reports the measured and estimated time with the respective error.
For all the applications, the relative error remains bellow 9.2\% with an average value of 7.6\%.



\ctable[caption={Kernel applications and estimated bandwidth.GMI-Global Memory Interconnect BCA-Burst Coalesced Aligned. BCNA-Burst Coalesced Non-Aligned. ACK-Burst Coalesced Write ACK. M- Measured. E- Estimated}, 
label=tab:applications, doinside=\small
]{llcS[table-format=3.1]S[table-format=3.1]c}{}{
\FL
Kernel & GMI  & $\#lsu$  & {M.Time} & {E.Time} & Error\\
         &     &         & [ms]   &  [ms]  & [\%]
\ML

Dot\cite{DeMatteis2019}             & BCA           & 3  &  60.2     &   64.5    & 7.3\NN
FFT-1D \cite{IntelQuartusPrime19}   & BCA           & 2  &  9.5      &   8.8     & 7.3   \NN  
nn\cite{Zohouri2016}                & BCA           & 2  &  157.5    &   172.1   & 9.2   \NN      
ROT\cite{DeMatteis2019}             & BCA           & 4  &  92.7     &   86.1    & 7.2  \NN
VectorAdd \cite{IntelQuartusPrime19}  & BCA         & 3  &  33.3     &   33.2    & 5.1  \NN  
VectorAdd\small{$\delta=2$}& BCA         & 3  &  67.9     &   63.0    & 6.5  \NN  
Hotspot\cite{Zohouri2016}           & BCNA          & 3  &  9.7      &   8.8     & 8.7   \NN
Pathfinder\cite{Zohouri2016}        & BCNA          & 3  &  275.9    &   254.0   & 7.9    \NN          
WM \cite{Vivado2017}                & BCNA          & 2  &  59.8     &   55.8    & 6.6    \NN
NW\cite{Zohouri2016}                & ACK          & 4  &  1.4      &   1.4    &  4.0
\LL
}

\subsection{Comparison with other models}

To compare the proposed model with two state-of-the-art models: Wang and HLScope+~\cite{Wang2016,Choi2017}, we have manually computed their estimations for the microbenchmarks, with $f=16$, and for the vectorAdd application. Unfortunately, the comparison of more applications is unfeasible because their dynamic profiling tools feeding the models are not publicly available. The tests are run with two BSPs with different DRAM frequency, 1866 and 2666 MHz.

In all, but one case, $\mu$b BCA, the error of this work is lower than that of Wang and HLSCope+ as Table~\ref{tab:compare} shows. Comparing the maximum error of each model, this proposal is up to 400 and 5$\times$ more accurate than Wang and HLScope+, respectively. 



\ctable[caption={Execution time estimated error; $\mu$b, BCA, BCN, and ACK refers to microbenchmark, burst coalesced aligned, burst coalesced non-aligned, and burst coalesced Write Ack, respectively.},
label=tab:compare,pos = ht, doinside=\small
]{lcS[table-format=3.1]S[table-format=3.1]S[table-format=3.1]}{}{
\FL
Benchmark & $\#lsu$     & {Wang}    & {HLScope+} & {This work}\NN
&             &     [\%]  & [\%]       &  \multicolumn{1}{c}{[\%]}
\ML
DDR4-1866
\ML
$\mu$b BCA & 1    & 17.3     & 12.7 & 5.6 \NN  
$\mu$b BCA & 4    & 0.3      & 10.6 & 4.4 \NN  
$\mu$b BCN & 3  & {-}       & 71.1 & 4.0  \NN  
$\mu$b ACK & 2     & 8049.9   & 63.2 & 27.9 \NN  
VectorAdd        & 3     & 19.3     & 21.0 & 5.1 
\ML
DDR4-2666
\ML
$\mu$b BCA & 1   & 69.6    & 57.8 & 4.7 \NN  
$\mu$b BCA & 4    & 37.8    & 19.6  & 5.8 \NN  
$\mu$b BCNA & 3  & {-}       & 137.9 & 8.7 \NN  
$\mu$b ACK & 2   & 11279.4 & 47.6 &  8.8 \NN  
VectorAdd       &  3   & 67.9    & 63.3  & 1.0
\LL
}

In Wang's case, the errors come from an incomplete support of all LSU modifiers and by not fully including the memory features (bandwidth, frequency, row misses, \ldots), which this work does. On the other hand, in HLScope+, for memory bound applications, the estimation is primary affected by DRAM bandwidth. Also, HLScope+ requires a board characterization to compute the controller overhead ($Tco$) \cite{ONeal2018}. This subsection uses $Tco=2.5ns$ for $\#lsu>3$, and $Tco=0ns$ in other cases.

In addition, please note that Wang and HLScope+ do not adapt well to changes in memory, contrary to this work that is able to take them into account.



\section{Related Work}
\label{sec:BACK}

For many HPC applications running on FPGAs, memory is the main bottleneck
~\cite{Nabi2016,DaSilva2014}, contrary to these proposals that only focus on continuous and stride patterns, this work covers all possible patterns.


In the race to improve productivity with HLS in FPGA, performance modeling is a requirement. 
These models can be classified between those based on dynamic profiling~\cite{Zhong2016,Makrani2019,Zhao2019,Zhao2017,Choi2017} an those based on static analysis~\cite{Wang2016}.
Some predictive models have been compared for FPGA by~\cite{ONeal2018}, where only \cite{Choi2017} covers external memory with the drawback that it requires on board characterization, and the memory interface is not portable to other boards~\cite{ONeal2018}. In the same way, works, as~\cite{Wang2016} for OpenCL, have a coarse grain model that shows inaccuracies in the memory behavior and requires non-traditional HLS flow, some of the limitations of this work has been detected by~\cite{Liang2018}. Additionally, Liang \emph{et al.}~\cite{Liang2018} improve models covering memory access patterns with a short CPU/GPU execution, but as some comparison shows, the memory controller makes difference in the performance~\cite{Zhang2016,Cong2918,Nabi2018}, moreover CPU/GPU devices have a more sophisticated memory hierarchy that can hide DRAM latency. In ~\cite{Waidyasooriya2017,Zohouri2018} they intent to guide programmers with analytical analysis of kernel, but the memory assumptions are limited without considering the memory interconnection, covered by our proposal.

\section{Conclusions}
\label{sec:CON}

Compilation time represents an adoption barrier for HLS  
with FPGAs. This paper proposes an analytical model, estimating the execution time of memory bound applications accurately, so programmers can quickly guess the code performance without the time consuming synthesis. 
The model stems from a detailed study of the generated RTL-code, instantiated IPs, and FPGA architecture and 
condenses the factors causing the memory delay for all possible types of memory accesses, vectorization factors, strides, \ldots without loss in flexibility. Even, it supports changes in the board support package such as DRAM speed. By focusing on memory bound applications, the model does not depend on the kernel pipeline and can be easily plugged into existing compute bound oriented models and HLS tools. The model is publicly available, so it can be easily extended if required. 

Our proposal has been carefully validated, and the obtained results show its accuracy predicting execution time. For 9 representative memory bound applications, the model error remains below 9\%, and, compared with two other state-of-the-art model, it reduces the error by at least 2$\times$. Our future work aims to integrate such models into scheduling policies of heterogeneous systems, where predicting performance before launching a kernel can make a difference for achieving a higher performance and energy efficiency.

\bibliographystyle{IEEEtran}
\bibliography{IEEEabrv,main.bib}

\begin{thebibliography}{10}
\providecommand{\url}[1]{#1}
\csname url@samestyle\endcsname
\providecommand{\newblock}{\relax}
\providecommand{\bibinfo}[2]{#2}
\providecommand{\BIBentrySTDinterwordspacing}{\spaceskip=0pt\relax}
\providecommand{\BIBentryALTinterwordstretchfactor}{4}
\providecommand{\BIBentryALTinterwordspacing}{\spaceskip=\fontdimen2\font plus
\BIBentryALTinterwordstretchfactor\fontdimen3\font minus
  \fontdimen4\font\relax}
\providecommand{\BIBforeignlanguage}[2]{{%
\expandafter\ifx\csname l@#1\endcsname\relax
\typeout{** WARNING: IEEEtran.bst: No hyphenation pattern has been}%
\typeout{** loaded for the language `#1'. Using the pattern for}%
\typeout{** the default language instead.}%
\else
\language=\csname l@#1\endcsname
\fi
#2}}
\providecommand{\BIBdecl}{\relax}
\BIBdecl

\bibitem{Waidyasooriya2017}
H.~M. Waidyasooriya, M.~Hariyama, and K.~Uchiyama, \emph{{Design of FPGA-based
  computing systems with openCL}}, 2017.

\bibitem{Gautier2016}
Q.~{Gautier}, A.~{Althoff}, {Pingfan Meng}, and R.~{Kastner}, ``Spector: An
  opencl fpga benchmark suite,'' in \emph{2016 International Conference on
  Field-Programmable Technology (FPT)}, 2016, pp. 141--148.

\bibitem{Liang2018}
Y.~{Liang}, S.~{Wang}, and W.~{Zhang}, ``Flexcl: A model of performance and
  power for opencl workloads on fpgas,'' \emph{IEEE Transactions on Computers},
  vol.~67, no.~12, pp. 1750--1764, 2018.

\bibitem{Verma2016}
A.~Verma, A.~E. Helal, K.~Krommydas, and W.-c. Feng, ``{Accelerating Workloads
  on FPGAs via OpenCL: A Case Study with OpenDwarfs},'' \emph{Computer Science
  Technical Reports}, no. Section IV, pp. 1----9, 2016.

\bibitem{Zohouri2016}
H.~R. Zohouri, N.~Maruyamay, A.~Smith, S.~Matsuoka, and M.~Matsuda,
  ``{Evaluating and optimizing OpenCL kernels for high performance computing
  with FPGAs},'' \emph{International Conference for High Performance Computing,
  Networking, Storage and Analysis, SC}, vol. 2016, no. November, p.~35, 2016.

\bibitem{Wang2016}
Z.~{Wang}, B.~{He}, W.~{Zhang}, and S.~{Jiang}, ``A performance analysis
  framework for optimizing opencl applications on fpgas,'' in \emph{2016 IEEE
  International Symposium on High Performance Computer Architecture (HPCA)},
  2016, pp. 114--125.

\bibitem{Choi2017}
Y.~K. Choi, P.~Zhang, P.~Li, and J.~Cong, ``{HLScope+,: Fast and accurate
  performance estimation for FPGA HLS},'' vol. 2017-November, 2017, pp.
  691--698.

\bibitem{Trimberger2015}
S.~M. {Trimberger}, ``Three ages of fpgas: A retrospective on the first thirty
  years of fpga technology,'' \emph{Proceedings of the IEEE}, vol. 103, no.~3,
  pp. 318--331, 2015.

\bibitem{IntelEMI_IP2019}
{Intel}, ``{External Memory Interface Handbook Volume 3: Reference Material},''
  2017.

\bibitem{IntelQuartusPrime19}
{Intel}, ``{Intel FPGA SDK for OpenCL Pro Edition: Getting Started Guide
  19.1},'' 2019.

\bibitem{Zheng2010}
H.~{Zheng} and Z.~{Zhu}, ``Power and performance trade-offs in contemporary
  dram system designs for multicore processors,'' \emph{IEEE Transactions on
  Computers}, vol.~59, no.~8, pp. 1033--1046, 2010.

\bibitem{IntelStratix}
{Intel}, ``{Intel{\textregistered} Stratix{\textregistered} 10 TX Product
  Table},'' 2019.

\bibitem{IntelAgilex}
{Intel}, ``{Intel{\textregistered} Agilex{\textregistered} I-Series SoC FPGA
  Product Table},'' 2019.

\bibitem{IntelEMIF2019}
{Intel}, ``{External Memory Interfaces Intel {\textregistered} Stratix
  {\textregistered} 10 FPGA IP User Guide},'' 2019.

\bibitem{Wang2014}
W.~{Wang}, T.~{Dey}, J.~W. {Davidson}, and M.~L. {Soffa}, ``Dramon: Predicting
  memory bandwidth usage of multi-threaded programs with high accuracy and low
  overhead,'' in \emph{2014 IEEE 20th International Symposium on High
  Performance Computer Architecture (HPCA)}, 2014, pp. 380--391.

\bibitem{DeMatteis2019}
T.~D. Matteis, J.~de~Fine~Licht, and T.~Hoefler, ``{FBLAS: Streaming Linear
  Algebra on FPGA},'' \emph{CoRR}, vol. abs/1907.07929, 2019.

\bibitem{Vivado2017}
\BIBentryALTinterwordspacing
X.~Vivado, ``{Vivado Design Suite User Guide: High-Level Synthesis},'' vol.
  901, pp. 1--120, 2017. [Online]. Available:
  \url{www.xilinx.com/products/design-tools/software-zone/sdaccel.html}
\BIBentrySTDinterwordspacing

\bibitem{ONeal2018}
K.~{O'Neal} and P.~{Brisk}, ``Predictive modeling for cpu, gpu, and fpga
  performance and power consumption: A survey,'' in \emph{2018 IEEE Computer
  Society Annual Symposium on VLSI (ISVLSI)}, 2018, pp. 763--768.

\bibitem{Nabi2016}
S.~W. Nabi and W.~Vanderbauwhede, ``{FPGA design space exploration for
  scientific HPC applications using a fast and accurate cost model based on
  roofline analysis},'' \emph{Journal of Parallel and Distributed Computing},
  pp. 1--13, 2016.

\bibitem{DaSilva2014}
B.~{Da Silva}, A.~Braeken, E.~H. D'Hollander, and A.~Touhafi, ``{Performance
  and resource modeling for FPGAs using high-level synthesis tools},'' in
  \emph{Advances in Parallel Computing}, vol.~25.\hskip 1em plus 0.5em minus
  0.4em\relax IOS Press BV, 2014, pp. 523--531.

\bibitem{Zhong2016}
G.~{Zhong}, A.~{Prakash}, Y.~{Liang}, T.~{Mitra}, and S.~{Niar},
  ``Lin-analyzer: A high-level performance analysis tool for fpga-based
  accelerators,'' in \emph{2016 53nd ACM/EDAC/IEEE Design Automation Conference
  (DAC)}, 2016, pp. 1--6.

\bibitem{Makrani2019}
H.~{Mohammadi Makrani} and et~al., ``Pyramid: Machine learning framework to
  estimate the optimal timing and resource usage of a high-level synthesis
  design,'' in \emph{2019 29th International Conference on Field Programmable
  Logic and Applications (FPL)}, 2019, pp. 397--403.

\bibitem{Zhao2019}
J.~{Zhao}, L.~{Feng}, S.~{Sinha}, W.~{Zhang}, Y.~{Liang}, and B.~{He},
  ``Performance modeling and directives optimization for high level synthesis
  on fpga,'' \emph{IEEE Transactions on Computer-Aided Design of Integrated
  Circuits and Systems}, pp. 1--1, 2019.

\bibitem{Zhao2017}
J.~{Zhao.}, L.~{Feng}, S.~{Sinha}, W.~{Zhang}, Y.~{Liang}, and B.~{He},
  ``Comba: A comprehensive model-based analysis framework for high level
  synthesis of real applications,'' \emph{2017 IEEE/ACM International
  Conference on Computer-Aided Design (ICCAD)}, pp. 430--437, Nov 2017.

\bibitem{Zhang2016}
C.~{Zhang}, {Zhenman Fang}, {Peipei Zhou}, {Peichen Pan}, and {Jason Cong},
  ``Caffeine: Towards uniformed representation and acceleration for deep
  convolutional neural networks,'' in \emph{2016 IEEE/ACM International
  Conference on Computer-Aided Design (ICCAD)}, Nov 2016, pp. 1--8.

\bibitem{Cong2918}
J.~{Cong}, Z.~{Fang}, M.~{Lo}, H.~{Wang}, J.~{Xu}, and S.~{Zhang},
  ``Understanding performance differences of fpgas and gpus,'' in \emph{2018
  IEEE 26th Annual International Symposium on Field-Programmable Custom
  Computing Machines (FCCM)}, April 2018, pp. 93--96.

\bibitem{Nabi2018}
S.~W. {Nabi} and W.~{Vanderbauwhede}, ``Mp-stream: A memory performance
  benchmark for design space exploration on heterogeneous hpc devices,'' in
  \emph{2018 IEEE International Parallel and Distributed Processing Symposium
  Workshops (IPDPSW)}, 5 2018, pp. 194--197.

\bibitem{Zohouri2018}
H.~R. Zohouri, A.~Podobas, and S.~Matsuoka, ``Combined spatial and temporal
  blocking for high-performance stencil computation on fpgas using opencl,'' in
  \emph{Proceedings of the 2018 ACM/SIGDA International Symposium on
  Field-Programmable Gate Arrays}, ser. FPGA ’18.\hskip 1em plus 0.5em minus
  0.4em\relax New York, NY, USA: Association for Computing Machinery, 2018, p.
  153–162.

\end{thebibliography}

\end{document}